\documentclass[letterpaper,12pt]{article}
\usepackage[utf8]{inputenc}
\usepackage[margin=1in]{geometry}
\usepackage{amsmath}
\usepackage{setspace}
\usepackage{indentfirst}
\usepackage{multirow,multicol}
\usepackage{float}
\usepackage[symbol]{footmisc}
\usepackage{natbib}
\usepackage{multirow,array}
\usepackage{booktabs,tabularx}
\usepackage[input-decimal-markers=.]{siunitx}
\usepackage{lipsum}
\usepackage{caption}
\usepackage{graphicx}
\usepackage[utf8]{inputenc}
\usepackage{adjustbox}
\usepackage[flushleft]{threeparttable}
\usepackage{fancyhdr}
\usepackage{placeins}
\usepackage{rotating}
\usepackage{bbm}
\usepackage{amsfonts}
\usepackage{hyperref}
\usepackage{booktabs,siunitx}
\usepackage{authblk}
\title{Demand and consumer surplus in the payday-loan market: Evidence from British Columbia}
\author[1]{Qi Zhang\thanks{Corresponding author: qitim.zhang@ucalgary.ca}}
\author[1]{Amity Quinn}
\affil[1]{ University of Calgary, Calgary, AB, Canada}

\date{\today }

\begin{document}
\maketitle
\vspace{-1.25em}

\begin{abstract}\noindent
This study examines how interest-rate caps affect the demand for payday loans, using aggregate data from British Columbia from 2012 to 2019, during which the province’s maximum fee was reduced from \$23 to \$17 and then to \$15 per \$100 borrowed. Estimating a linear demand function via OLS, we find that lowering interest-rate caps significantly increases loan demand. We estimate that the \$8 decrease from \$23 to \$15 raised annual consumer surplus by roughly \$28.6 million (2012 CAD). A further reduction to \$14, starting in January 2025, is projected to add an additional \$3.9 million per year. These results suggest that stricter interest-rate caps within a certain range can yield substantial consumer welfare gains, with no evidence of restricted access to payday loans due to lender exit from the market.
\end{abstract}
\vspace{-0.5em}

\section{Introduction}
Payday loans are typically extended to borrowers with limited liquidity. The high cost of this credit can lead to repeat borrowing and persistent debt cycles (\cite{stegman2007payday}). Users of these products in North America are disproportionately young and from marginalized communities; often women with have low or volatile incomes, precarious employment, and thin or impaired credit histories (\cite{buckland2018socio,apgar2006subprime,visano2008different,chen2022national,spotton2018mainstream}). To reduce the risk of exploitative practices and protect vulnerable consumers, most jurisdictions in the United States and Canada have enacted payday lending regulations. One common measure is a cap on borrowing costs, aimed at curbing the harmful features of payday lending (\cite{kaufman2013payday,dilay2018payday}).

Payday lenders typically price at or near the prevailing cap (\cite{avery2011payday, robinson2018business}). For example, a 15\% cap (i.e., a \$15 fee per \$100 borrowed for a two-week loan) translates into an annual percentage rate (APR) of approximately 391\%. Where market power exists, price ceilings may enhance efficiency. In addition, interest rate caps can serve as a rudimentary form of social insurance by reallocating surplus from lower-marginal-utility lenders to higher-marginal-utility borrowers (\cite{glaeser1998neither}).

Studies on interest-rate caps for mainstream consumer loans, which typically have APRs between 6\% and 50\%, show that increasing interest-rate caps raises the likelihood that credit-constrained consumers can obtain a loan, whereas lowering the cap may restrict access and potentially harm these borrowers (\cite{rigbi2013effects,madeira2019impact}).  However, the population targeted by payday loans is very different from that of mainstream consumer loans, and there is little existing evidence examining how interest rates affect the demand for payday loans and the resulting consumer surplus. Only one earlier study exploited transaction-level data to examine the reduction of the maximum fee per \$100 borrowed from \$15 to \$10 in Rhode Island and estimated an annual consumer-surplus gain of approximately \$3.4 million (USD) (\cite{fekrazad2020impacts}). In this study, we exploit multiple policy changes in British Columbia—which has a payday loan market roughly five times larger than Rhode Island’s—to analyze cap reductions between 2012 and 2019 that lowered the maximum fee per \$100 borrowed from \$23 to \$17, and subsequently to \$15. Using provincial-level aggregate data on borrowing costs and total loan volume, we estimate a linear demand curve for payday loans and calculate the associated welfare effects. We estimate that the \$8 reduction from \$23 to \$15 increased consumer surplus by approximately \$28.6 million per year (2012 CAD). A more recent policy, effective January 2025, further reduces the cap from \$15 to \$14 per \$100, and we estimate that this change will generate an additional \$3.9 million (2012 CAD) in annual consumer surplus.

\section{Data}
The payday-loan industry in British Columbia is licensed and regulated by Consumer Protection BC, which compiles annual operational statistics from all licensed lenders. The available dataset includes each year’s average borrowing cost (the fee charged per \$100 loan on a two-week term), total loan volume in nominal dollars, and total loan volume in real 2012 CAD (adjusted for inflation). Our sample covers 2012 through 2023, and summary statistics are provided in Table 1 (loan volumes are rounded to the nearest \$0.1 million). Over this period, the regulated maximum fee per \$100 was progressively lowered from \$23 (2012–2016) to \$17 in 2017, and to \$15 from 2018 onward. In practice, lenders priced loans at or very near the cap in each year. The reported annual average borrowing cost can slightly exceed the nominal cap in some years due to reporting cycle misalignment. 

It is also worth noting that the number of physical storefront lenders began declining in 2015, rebounded slightly in 2018 and 2019, and has continued to decline since then, while online payday lending in British Columbia began expanding in 2020.


\begin{table}[htbp]\centering
\caption{British Columbia payday-loan market statistics, 2012-2023}
\label{tab:bc_summary}
\begin{tabular}{l c c c c c }
\toprule
Year&Borrowing cost&Annual loan volume&Total loan volume& \# of locations &\# of locations  \\
& (per CAD 100)& (nominal CAD) & (2012 CAD) & (H.O. +Branches) & (Online only) \\
\midrule
2012 & 21.50 & 318.1 Mil & 318.1 Mil &274&- \\
2013 & 21.70 & 351.4 Mil & 344.4 Mil&275&-  \\
2014 & 21.90 & 385.3 Mil & 370.3 Mil&274&- \\
2015 & 21.70 & 340.9 Mil & 321.2 Mil &226&- \\
2016 & 21.70 & 369.7 Mil & 341.5 Mil &209&- \\
2017 & 19.30 & 397.3 Mil & 359.8 Mil &192&-  \\
2018 & 16.82 & 416.1 Mil & 369.5 Mil &202&- \\
2019 & 15.30 & 441.5 Mil & 384.4 Mil&204&-  \\
2020 & 14.92 & 390.7 Mil & 333.5 Mil & 194&14 \\
2021 & 14.69 & 271.5 Mil & 227.2 Mil&187 &15  \\
2022 & 14.85 & 305.7 Mil & 250.8 Mil &173&21 \\
2023 & 14.92 & 327.2 Mil & 263.2 Mil&170&21  \\
\bottomrule
\end{tabular}
\par\smallskip\footnotesize\emph{Notes:} Borrowing cost is the dollar fee per CAD 100 borrowed on 14-day terms. Loan volumes are annual totals and are  rounded to the nearest 0.1 million CAD. Real dollars are deflated to 2012 CAD. 
\end{table}

\section{Model}
\subsection{Demand of payday loans}
As in \cite{fekrazad2020impacts}, we assume that borrowers face a linear inverse demand curve for payday loans, given by:
\begin{equation}
Q=a-bp,\qquad b>0,
\label{eq:dem}
\end{equation}
where $p$ is the borrowing cost per CAD 100 (the fee per \$100 loan for a two-week term) and $Q$ is the total annual real loan volume (in year 2012 values). Under a binding interest rate cap, $p$ is effectively set by policy and $Q$ is determined by demand at that price.

\subsection{Estimation of demand}
To empirically estimate the demand curve, we focus on the period before 2020. We exclude observations from 2020–2023 because the Canadian government’s temporary COVID-19 income-support programs during those years materially altered consumers’ short-term credit needs. These developments could confound the price–quantity relationship in the later data. Given that we have only annual observations, adding additional control variables or policy dummies is not feasible (doing so would exhaust our degrees of freedom and obscure identification). We therefore restrict the sample to 2012–2019 and run a simple OLS regression of annual real loan volume (2012 CAD) on the actual borrowing cost per \$100.

\begin{table}[htbp]\centering
\caption{Demand estimation: real loan volume on borrowing cost (2012--2019)}
\label{tab:demand_reg_bc}
\begin{threeparttable}
\begin{tabular}{lc}
\toprule
\multicolumn{2}{l}{\textbf{DV: Real loan volume (2012 \$)}}\\
\midrule
Borrowing cost (per \textdollar {}100) & -6,621,526**\\
 & (2,618,230)\\[2pt]
Constant & 483,509,000*** \\
 & \multicolumn{1}{c}{(52,700,000)} \\
\midrule
Observations & 8 \\
R-squared & 0.5154 \\
F-statistic & 6.38 \\
Prob $>$ F & 0.0449 \\
\bottomrule
\end{tabular}
\begin{tablenotes}[flushleft]\footnotesize
\item Notes: OLS regression of annual real loan volume (2012 CAD) on the borrowing cost per \$100, restricted to years with fee cap above the 2020 change (2012--2019; $N=8$). Robust standard errors in parentheses. $^{*}p<0.10$, $^{**}p<0.05$, $^{***}p<0.01$.
\item 
\end{tablenotes}
\end{threeparttable}
\end{table}

Although this yields only eight data points, prior research suggests that even a small sample can produce informative estimates when the underlying relationship is strong and stable (\cite{jenkins2020solution}). Consistent with theory, the estimated slope coefficient is negative and statistically significant: $\hat{\beta} = -6.61$ million CAD (SE = 2.62 million). This implies that a \$1 increase in the fee per \$100 is associated with about a 6.61 million CAD decrease in annual loan volume. The estimated intercept is $\hat{\alpha} = 483.5$ million CAD (SE = 52.7 million). These OLS results are reported in Table 2.

\section{Consumer surplus gains from lower interest rate caps}
We quantify consumer-surplus (CS) gains from two cap reductions: (i) lowering the cap from \$23 to \$15 per \$100 borrowed, and (ii) lowering it further from \$15 to \$14 per \$100 (implemented Canada-wide in January 2025). Using the estimated demand curve $\hat{Q}(p)=\hat{a}-\hat{b}p$ with $\hat{a}=483.509$ and $\hat{b}=6.621$ (where $Q$ is in millions of 2012 CAD and $p$ in dollars per \$100), predicted volumes are
\[
Q_{23} = \hat{Q}(23) = 331.226~\text{million},\quad
Q_{15} = \hat{Q}(15) = 384.194~\text{million},\quad
Q_{14} = \hat{Q}(14) = 390.815~\text{million}.
\]
The CS change from reducing the cap from $p_1$ to $p_2$ ($p_2<p_1$) equals the area under demand between $p_1$ and $p_2$:
\begin{equation}
\Delta CS(p_1\to p_2)
= \frac{1}{100}\int_{p_2}^{p_1}\hat{Q}(p)\,dp
= \frac{1}{100}\Big[\hat{a}(p_1-p_2)-\frac{\hat{b}}{2}\big(p_1^2-p_2^2\big)\Big],
\label{eq:cs}
\end{equation}
where the factor $1/100$ converts the fee (quoted per \$100) to a per \$1 price for integration. Plugging in $\hat{a}$ and $\hat{b}$, we obtain $\Delta CS_{23\to 15}\approx 28.62$ million (2012 CAD) and $\Delta CS_{15\to 14}\approx 3.88$ million (2012 CAD) per year. The latter figure is the projected annual benefit from the new cap introduced in 2025.

\section{Robustness Check}

To further ensure the reliability of our primary findings, we estimate a quadratic demand curve using 2012--2019 data, modeling real annual payday loan volume $Q(p)$ (in 2012 dollars) as a function of the cost of borrowing $p$ (in dollars per \$100):
\[
Q(p) = a - b\,p + c\,p^2, b>0,
\]
where the estimated coefficients from OLS are:
\[
\hat a = 671{,}000{,}000, \quad \hat b = 26{,}900{,}000, \quad \hat c = 539{,}434.
\]
This functional form allows for curvature in borrower response to price, with a slight flattening at higher prices.

Evaluating $\hat Q(p)$ at the relevant cap levels yields:
$\hat Q(23) = 337.936$ million, $\hat Q(15)=388.872 $ million, $\hat Q(14)=400.130$ million. 

The change in consumer surplus (CS) when price falls from $p_1$ to $p_2$ is:
\begin{equation}
\Delta CS(p_1\to p_2)
= \frac{1}{100}\int_{p_2}^{p_1}\hat{Q}(p)\,dp
= \frac{1}{100}\Big[\hat{a}(p_1-p_2)-\frac{\hat{b}}{2}\big(p_1^2-p_2^2\big)+\frac{\hat{c}}{3}\big(p_1^3-p_2^3\big)\Big].
\label{eq:cs2}
\end{equation}
 Plugging in $\hat{a}$ and $\hat{b}$, we obtain $\Delta CS_{23\to 15}\approx 28.59$ million (2012 CAD) and $\Delta CS_{15\to 14}\approx 3.94$ million (2012 CAD) per year. These figures are very close to the estimates obtained from the linear demand specification.

\section{Conclusion}
Using a small set of annual aggregate data, we estimate a linear demand curve for payday loans and approximate the associated consumer surplus. Despite the coarse granularity of the data—and the necessarily approximate welfare calculations—the results are consistent and policy-relevant. In British Columbia, lowering the cap from \$23 to \$15 per \$100 borrowed coincided with a sizable increase in annual loan volume (from roughly \$331 million to \$384 million, 2012 CAD) and an estimated \$28.6 million gain in consumer surplus per year. A further cap reduction to \$14 per \$100 (implemented nationally in 2025) is projected to add about \$3.9 million annually. These findings are consistent with an earlier study that investigated a similar policy change in the United States, which suggests that tighter interest-rate ceilings can enhance borrower welfare without suppressing overall loan usage (\cite{fekrazad2020impacts}).

The contraction of storefront lenders alongside the rise of online providers points to heterogeneous cost structures: online firms may profitably extend credit at lower fees than brick-and-mortar lenders. Nevertheless, policymakers should consider potential supply-side responses when evaluating further cap reductions. Prior research suggests that lowering the interest-rate cap from \$15 to \$10 per \$100 borrowed may have contributed to lender exit and reduced consumer access in Oregon (\cite{zinman2010restricting}). However, these findings should be interpreted with caution, as the average payday-loan default rate in Oregon—approximately 10 to 12 percent—is substantially higher than the 2 to 5 percent range observed in British Columbia. In contrast, a more recent study examining the same policy change in Rhode Island, where default rates are similar to those in British Columbia, found no evidence of reduced access resulting from the cap reduction (\cite{fekrazad2020impacts}).

Our main findings are based on only a few aggregate data points from a single province in Canada, and much more can be learned from future research using aggregate data from multiple jurisdictions and micro-level data to examine heterogeneity across borrower segments, default risk, and lender margins, as well as to assess substitution toward other forms of credit. Overall, our evidence supports carefully calibrated caps within a certain range as a viable consumer-protection tool in high-cost credit markets—provided that regulators continue to monitor market composition and credit access as the industry evolves.

\section*{Funding}
No funding was received for this work.
\section*{Disclosure statement}
No potential conflict of interest was reported by the authors.


\newpage 
\singlespacing
\bibliography{reference.bib}
\bibliographystyle{ts.bst}
\end{document}